\documentclass[aps,prl,twocolumn,superscriptaddress,showpacs]{revtex4}

\usepackage{graphicx} % uncomment this line to include the graphicx package
\usepackage{amssymb} % for the square root function

\usepackage{amsmath} %follow EK
\usepackage{dcolumn} %follow EK
\usepackage{bm}      %follow EK

\begin{document}

%\preprint{}

\title{The effect of $^3$He impurities on the nonclassical response to
oscillation of solid $^4$He}



\author{E. Kim}
\altaffiliation{Permanent address: Physics Department, Korea
Advanced Institute of Science and Technology (KAIST), 373-1
Guseong-dong, Yuseong-gu, Daejeon 305-701, Korea}
\affiliation{Department of Physics, The Pennsylvania State
University, University Park, Pennsylvania 16802 USA}
\author{J. S. Xia}
\affiliation{Department of Physics, University of Florida,
Gainesville, Florida 32611 USA} \affiliation{National High Magnetic
Field Laboratory (NHMFL), Tallahassee, Florida 32310 USA}
\author{J. T. West}
\email{jtw11@psu.edu}
\author{X. Lin}
\author{A. C. Clark}
\author{M. H. W. Chan}
\affiliation{Department of Physics, The
Pennsylvania State University, University Park, Pennsylvania 16802
USA}

%\thanks{}

\date{\today}

\begin{abstract}

We have investigated the influence of impurities on the possible
supersolid transition by systematically enriching isotopically-pure
$^4$He ($<$ 1 ppb of $^3$He) with $^3$He. The onset of nonclassical
rotational inertia is broadened and shifts monotonically to higher
temperature with increasing $^3$He concentration, suggesting that
the phenomenon is correlated to the condensation of $^3$He atoms
onto the dislocation network in solid $^4$He.

\end{abstract}

%\keywords{}

\maketitle

The observation of nonclassical rotational inertia (NCRI) in solid
helium was first reported in a torsional oscillator (TO) experiment
with the solid confined within porous Vycor glass \cite{KC1}. The
NCRI signal is measured as a drop in the resonant period $\tau_0$ of
the TO. An intriguing and perhaps counterintuitive result of this
experiment is the extreme sensitivity to $^3$He impurities in the
solid. When a minute concentration of $^3$He (\textit{x}$_3$
$\approx$ 10 ppm) was present the NCRI fraction (NCRIF) showed a
20\% decrease. NCRIF is defined by normalizing the apparent mass
decoupling in the low temperature limit by the total mass loading of
the solid helium sample. The onset temperature \textit{T$_O$} (the
point where NCRI becomes resolvable from the noise) was found to
increase from $\sim$175 mK to $\sim$300 mK. Increasing
\textit{x}$_3$ beyond the 10 ppm level continued to increase
\textit{T$_O$} and decrease NCRIF until, at just \textit{x}$_3$ =
0.1\%, the signal became undetectable.

NCRI has also been reported in TO measurements on bulk solid $^4$He
\cite{KC2,KC4,RR1,RR2,HK,MK,KS,clark}. All studies except one
\cite{clark} were carried out with commercially available,
ultra-high purity (UHP) $^4$He solid samples grown using the blocked
capillary (BC) method. In all of the data published to date, the
temperature dependence in UHP $^4$He (both in the bulk
\cite{KC2,KC4,RR1,RR2,HK,MK,KS,clark} and in porous media
\cite{KC1,KC3}) is qualitatively reproducible. With decreasing
temperature, NCRI gradually emerges from the background near
\textit{T$_O$} and then rapidly increases before reaching a constant
value below a point of saturation \textit{T$_S$}. These
characteristic temperatures typically fluctuate by a factor of two.
In contrast, NCRIF varies widely \cite{RR2}, from 0.03\% to at least
20\%. It has been suggested that this reflects the degree of
disorder in the solid, with higher quality crystals having a smaller
NCRIF. However, experimental evidence does not consistently support
this notion. For example, although superior growth techniques tend
to reduce the effect in a particular TO, NCRIF can still vary by a
factor of ten in large crystals grown at constant pressure in two
different cells \cite{clark}.

A different method to study the effects of disorder is to introduce
point defects into the crystal. In view of the sensitivity of the
phenomenon to $^3$He at the parts per million level \cite{KC1}, we
have carried out a systematic study in the bulk phase in which the
$^3$He concentration was varied from below 1 ppb up to 30 ppm. Two
different torsional oscillators were fabricated for this experiment.
Measurements of $^4$He samples with 1 ppb $\leq$ \textit{x}$_3$
$\leq$ 129 ppb were carried out at the high B/T facility of NHMFL at
the University of Florida employing a TO (TOF) with $\tau_0$ = 0.771
ms and mechanical quality factor \textit{Q} = 1 x 10$^6$. The
cylindrical sample space in TOF has a height, \textit{h} = 0.50 cm,
and a diameter, \textit{d} = 1.00 cm. The period increases by
$\Delta\tau_0$ = 3940 ns upon filling the cell with solid helium at
60 bar. Concentrations in the range, 70 ppb $\leq$ \textit{x}$_3$
$\leq$ 30 ppm, were studied at Penn State with another TO (TOP)
having $\tau_0$ = 1.277 ms, \textit{Q} = 5 x 10$^5$, \textit{h} =
0.64 cm, \textit{d} = 0.76 cm, and $\Delta\tau_0$ = 1170 ns.

The isotopically-pure gas came from the U.S. Bureau of Mines.
Similarly purified gas was analyzed \cite{1ppb} and found to have
\textit{x}$_3$ $<$ 1 ppb. There is some uncertainty in the exact
concentration of the UHP $^4$He. The precise value varies by the
source \cite{UHP} but is always less than 1 ppm. For calculating
mixture concentrations we use \textit{x}$_3$ = 300 ppb for the
commercial UHP gas and \textit{x}$_3$ = 1 ppb for the
isotopically-pure gas. Samples with \textit{x}$_3$ $<$ 300 ppb were
prepared by mixing appropriate amounts of the 1 ppb and 300 ppb
gases. Samples with \textit{x}$_3$ $>$ 300 ppb were prepared by
mixing 300 ppb gas with pure $^3$He. To avoid contamination from
residual $^4$He a new capillary and room temperature gas handling
system were constructed for the isotopically-pure samples.
Measurements in each cell started with samples of the lowest
\textit{x}$_3$, followed by studies of progressively higher
concentrations. The pressure of the solid is determined using an
\textit{in situ} resistive strain gauge (resolution $\approx$ 0.5
bar) attached directly to the outer wall of the torsion cell. The BC
method was used for all samples resulting in pressures of 60 $\pm$ 5
bar. All of the data were taken during warming scans with low
oscillation amplitudes, where the maximum linear speed
\textit{v$_{RIM}$} of the TO was near or below 10
$\mu$m$\,$s$^{-1}$.

\begin{figure}[t]
\includegraphics[width=1.0\columnwidth]{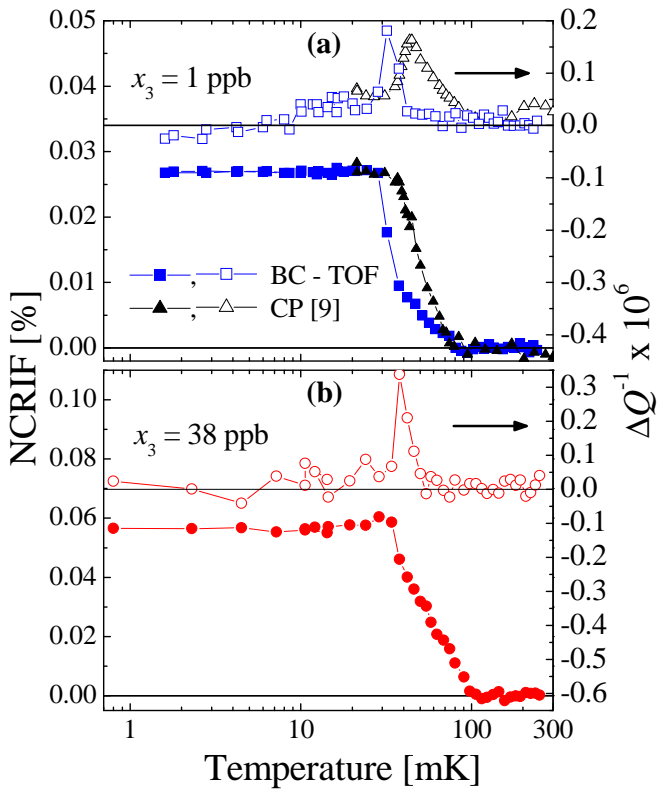}
\caption{\label{fig:one}(a) Comparison of 1 ppb samples from TOF and
Ref.~\cite{clark}. No additional temperature dependence is found
below \textit{T$_S$}. NCRIF and \textit{Q}$^{-1}$ data for the CP
sample \cite{clark} are divided by 12 and 5, respectively. (b)
Ultralow temperature scan of a 38 ppb sample from TOF.}
\end{figure}

\begin{figure}[t]
\includegraphics[width=1.0\columnwidth]{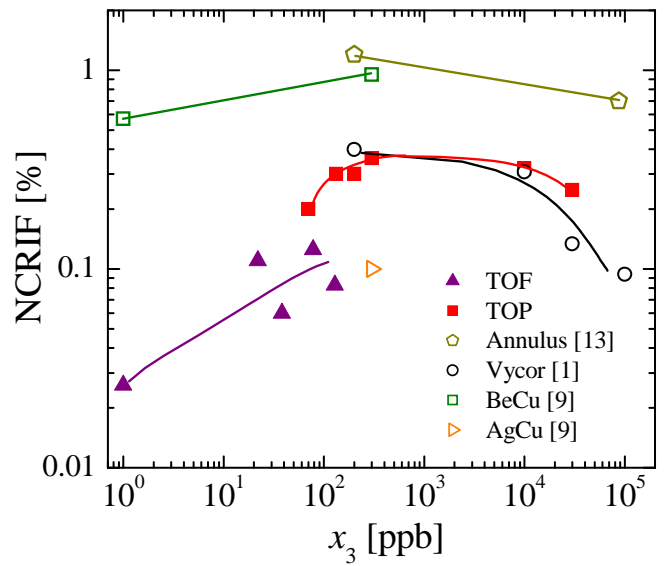}
\caption{\label{fig:two}$^3$He dependence of NCRIF for BC samples
obtained in this study and in other TO's from our laboratory. There
appears to be an optimal concentration of $\sim$1 ppm.}
\end{figure}

Fig.~\ref{fig:one}(a) shows NCRIF as a function of temperature for a
solid sample with \textit{x}$_3$ = 1 ppb. NCRI appears below
\textit{T$_O$} $\approx$ 80 mK and saturates below \textit{T$_S$} =
28 mK. A small dissipation peak is observed with a peak temperature,
\textit{T$_P$} = 32 mK. For \textit{T} $>$ \textit{T$_S$}, NCRIF
initially drops rapidly and then exhibits a much slower decay to
zero between 40 mK and 80 mK. Measurements down to 1 mK reveal no
additional features below \textit{T$_S$}. Similar behavior is seen
at \textit{x}$_3$ = 38 ppb [see Fig.~\ref{fig:one}(b)]. Data from a
sample (\textit{x}$_3$ = 1 ppb) grown at constant pressure (CP) in a
different cell \cite{clark} are also shown in Fig.~\ref{fig:one}(a).
Although the onset temperatures of the two samples are nearly
identical, the CP sample (expected to be of much higher quality than
the BC sample) shows a sharper transition near \textit{T$_O$}.

In Fig.~\ref{fig:two} we examine the \textit{x}$_3$ dependence of
NCRIF measured in the low temperature limit for both this study and
for other data obtained at Penn State. Despite the significant shift
in the magnitude of NCRIF from cell to cell, the consistent trend is
that NCRIF first increases with the impurity concentration and then
decreases with further $^3$He enrichment beyond $\sim$1 ppm.

\begin{figure}[t]
\includegraphics[width=1.0\columnwidth]{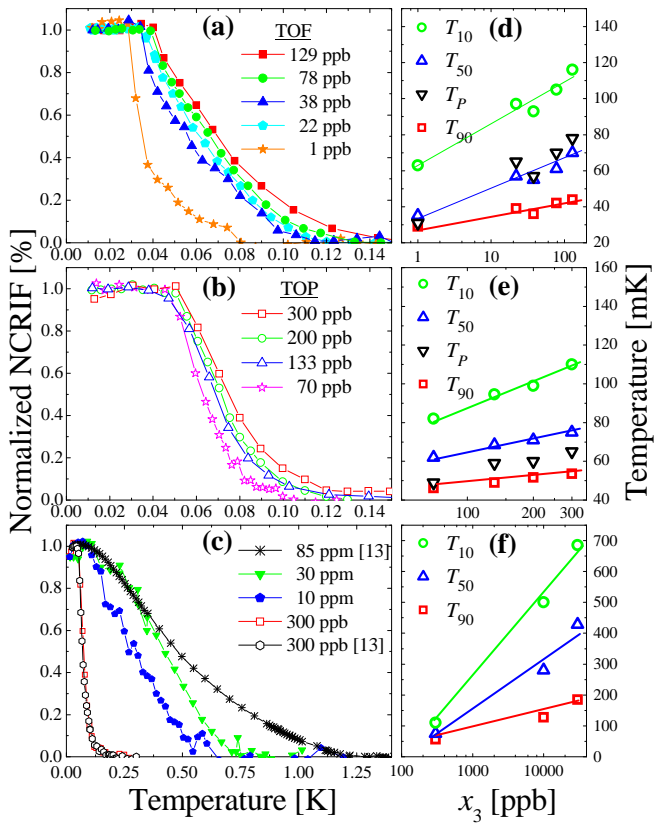}
\caption{\label{fig:three}Temperature dependence of NCRIF
(normalized by the low temperature limiting value) for different
\textit{x}$_3$ grown in (a) TOF and (b) TOP. (c) Temperature
dependence for \textit{x}$_3$ $\geq$ 300 ppb. Two traces obtained
with an annular cell \cite{thesis} are also shown. (d), (e), (f) The
right three panels show the \textit{x}$_3$ dependencies of the
characteristic temperatures, which were extracted from the adjacent
plots.}
\end{figure}

We compare several solid samples with different \textit{x}$_3$ in
Fig.~\ref{fig:three}. NCRIF is normalized to focus on the
temperature dependence. Figure~\ref{fig:three}(a) shows data from
TOF. The data presented in Figs.~\ref{fig:three}(b) and
\ref{fig:three}(c) are from TOP, with the exception of that obtained
\cite{thesis} with the same annular cell used in Refs.
\cite{KC2,KC4}. Both the NCRIF and \textit{Q}$^{-1}$ (not shown)
curves become increasingly broad and shift to higher temperature
with increasing $^3$He concentration. For samples of overlapping
\textit{x}$_3$, data from TOF and TOP are consistent.

\begin{figure}[t]
\includegraphics[width=1.0\columnwidth]{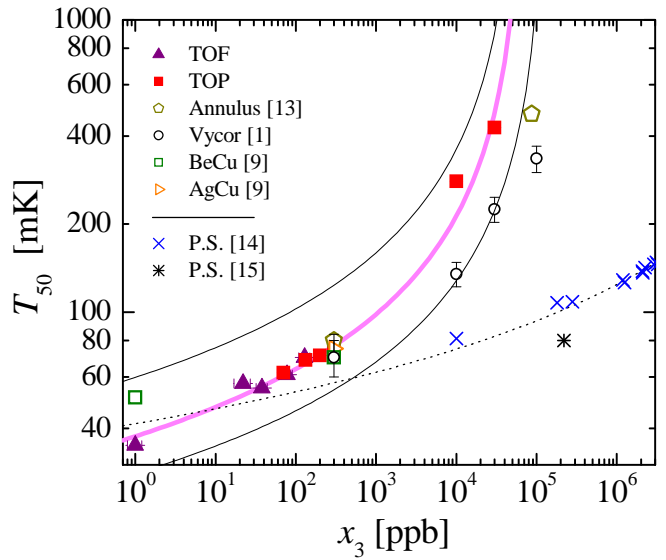}
\caption{\label{fig:four}\textit{T}$_{50}$ versus \textit{x}$_3$.
Relative to \textit{T}$_{50}$, values of \textit{T}$_{10}$ and
\textit{T}$_{90}$ (excluded for clarity) are shifted vertically
upward by $\sim$60\% and downward by $\sim$25\%, respectively. From
top to bottom, the three solid lines represent the condition,
\textit{T$_{IP}$} = \textit{T$_x$}, for \textit{x} = 10, 50, and 90.
To apply Eq.~(\ref{eq:one}) we assume that all samples have the same
\textit{L$_N$} (and $\Lambda$). The fitting parameters are listed in
Table~\ref{tab:table1}. The theoretical phase separation boundary
\cite{EB} (dotted line), anchored by pressure \cite{ganshin} and
ultrasound \cite{goodkind} measurements, is inconsistent with the TO
data.}
\end{figure}

Due to the asymptotic decay to zero of NCRI at high temperature and
the rounding near the low temperature saturation point, it is
difficult to precisely determine \textit{T$_O$} and \textit{T$_S$}.
To quantitatively compare the temperature dependence in
Fig.~\ref{fig:three} we have plotted the temperatures,
\textit{T$_x$} [see Figs.~\ref{fig:three}(d), \ref{fig:three}(e),
and \ref{fig:three}(f)], at which the normalized NCRIF is
\textit{x}\% (where \textit{x} = 10, 50, or 90) of its low
temperature limiting value. The locations of the dissipation peak
\textit{T$_P$} are also shown, revealing that at low $^3$He
concentrations \textit{T$_P$} $\approx$ \textit{T}$_{50}$. As
\textit{x}$_3$ increases so does the value of \textit{T$_P$}, but
less dramatically than \textit{T}$_{50}$. The broadening of the
dissipation peak with increasing \textit{x}$_3$ is such that no
well-defined peak is observable at the highest concentrations. We
have compiled the values of \textit{T}$_{50}$ from this and other
experiments into Fig.~\ref{fig:four}.

One nonsuperfluid mechanism that we have considered as an
explanation of the TO experiments is phase separation of the dilute
$^3$He-$^4$He mixtures. We show in Fig.~\ref{fig:four} the phase
separation boundary according to both experiments
\cite{ganshin,goodkind} and theory \cite{EB}. The discrepancy at
high \textit{x}$_3$ (even for \textit{T}$_{90}$) makes it clear that
our observations are not the result of phase separation. Although
the theoretical boundary at low \textit{x}$_3$ crosses the
datapoints from this study, there is no experimental evidence
\cite{goodkind,xi} of phase separation for \textit{x}$_3$ $<$ 27
ppm.

It has been proposed \cite{anderson} that the dependencies of NCRI
on temperature, velocity, and $^3$He are the result of a vortex
liquid phase, with the true supersolid transition occurring at lower
temperature. In this context, the high temperature tail reflects the
finite response time of vortices in the sample to oscillatory
motion. The broadening with increasing \textit{x}$_3$ (see
Fig.~\ref{fig:three}) is due to the slowed vortex motion when $^3$He
atoms attach to the normal cores and get dragged along with the
vortices. The accompanying dissipation peak signifies the matching
of the TO resonant frequency and the optimal rate at which vortices
can respond to changes in the velocity field. Thus, the
\textit{x}$_3$ dependence of \textit{T$_P$} (which is similar to
that of \textit{T$_x$}) may be a direct probe of the drag force
caused by $^3$He impurity atoms. The complete saturation of NCRIF
below \textit{T$_S$} (see Fig.~\ref{fig:one}) indicates that, if
there is a supersolid phase of $^4$He, the critical temperature
\textit{T$_C$} is either less than 1 mK or is such that
\textit{T$_S$} $<$ \textit{T$_C$} $<$ \textit{T$_O$}.

\begin{table}[b]
\caption{\label{tab:table1}Parameters for Eq.~(\ref{eq:one}) that
are associated with the three curves in Fig.~\ref{fig:four}.
Uncertainties in \textit{E$_B$} are 10\%, and in \textit{L$_N$} they
are 20\%. The dislocation densities were calculated by setting
\textit{L$_N$} $\approx$ \textit{L$_{IP}$} and
$\Lambda$\textit{L$_N$}$^2$ = 0.2.}
\begin{ruledtabular}
\begin{tabular}{cccc}
\textit{T$_x$} [K]&\textit{E$_B$} [K]&\textit{L$_{IP}$} [$\mu$m]&$\Lambda$ [10$^7$ cm$^{-2}$]\\
\hline
\textit{T}$_{10}$ & 0.66 & 1.7 & 0.7\\
\textit{T}$_{50}$ & 0.42 & 1.8 & 0.6\\
\textit{T}$_{90}$ & 0.33 & 1.3 & 1.2\\
\end{tabular}
\end{ruledtabular}
\end{table}

The similarities between vortices and dislocations have been
discussed recently \cite{umassclark}. Thus, another process that we
consider is the condensation (evaporation) of $^3$He atoms onto
(from) dislocation lines within the sample upon cooling (warming)
\cite{iwasa,PBD}. Dislocations in solid $^4$He form a random
three-dimensional network in which the lines intersect with one
another to form a vast number of nodes, which are essentially
immobile. In contrast, the line segments between nodes can move
readily in stress fields. When an oscillating stress field is
imposed, the dislocation segments vibrate with little or no damping
below 1 K \cite{wanner}. The network is characterized by the total
line density $\Lambda$ (total dislocation line length per unit
volume) and network loop length \textit{L$_N$} between nodes. From
ultrasound measurements of single crystals
\cite{iwasa,wanner,iwasa'}, it is usually found that 0.1 $<$
$\Lambda$\textit{L$_N$}$^2$ $<$ 0.2, where $\Lambda$ $\approx$ 1 x
10$^6$ cm$^{-2}$ and \textit{L$_N$} $\approx$ 5 $\mu$m. In crystals
grown with the BC method it is expected that $\Lambda$ will increase
and \textit{L$_N$} will decrease. At low temperature $^3$He atoms
bind to the dislocations with an energy \textit{E$_B$} and act as
additional pinning centers. There is a crossover from
network-pinning to impurity-pinning when the average distance
\textit{L$_{IP}$} between condensed $^3$He atoms becomes less than
\textit{L$_N$}. The \textit{x}$_3$-dependent temperature at which
this occurs can be obtained from the average pinning length
\cite{iwasa}, and is of the form
\begin{equation}\label{eq:one}T_{IP}=-2E_B\left(\ln{\left[\frac{x_3^2L_{IP}^3E_B}{4\mu b^6}\right]}\right)^{-1}.\end{equation}
Here, \textit{b} is the magnitude of the Burger's vector of a
dislocation and $\mu$ is the shear modulus of $^4$He.

In order to reveal the possible connection between $^3$He
impurity-pinning and the observed \textit{x}$_3$ dependence of NCRI,
we identify the crossover point with each of the characteristic
temperatures (i.e., \textit{T$_{IP}$} = \textit{T$_x$}) and fit the
measured \textit{T}$_{10}$, \textit{T}$_{50}$, and \textit{T}$_{90}$
by adjusting the parameters \textit{L$_{IP}$} and \textit{E$_B$}.
Figure~\ref{fig:four} demonstrates the accuracy of
Eq.~(\ref{eq:one}) in describing \textit{T}$_{50}$ versus
\textit{x}$_3$. The best fit parameters, which are consistent with
those found in the literature \cite{iwasa,PBD}, are listed in
Table~\ref{tab:table1}. If we fix \textit{E$_B$} = 0.42 K for all
three datasets, the curves calculated from Eq.~(\ref{eq:one})
deviate from the observed \textit{T}$_{10}$ and \textit{T}$_{90}$
for \textit{x}$_3$ $<$ 100 ppb. However, such a protocol results in
\textit{L$_{IP}$} that are longest for \textit{T}$_{10}$ and
shortest for \textit{T}$_{90}$, as expected \cite{iwasa}. The same
qualitative trends are observed in the Vycor data. However, the
above analysis is inappropriate due to the solid $^4$He morphology
\cite{knorr} and overall complexity of the system. If we naively
apply Eq.~(\ref{eq:one}) we get parameters similar to those in
Table~\ref{tab:table1} (eg., for \textit{T}$_{50}$ we get
\textit{L$_{IP}$} = 0.5 $\mu$m, which is much greater than the 7 nm
pore size).

The fact that the characteristic temperatures of NCRI can be
described by Eq.~(\ref{eq:one}) indicates that the observed
\textit{x}$_3$ dependence is correlated with the impurity-pinning of
dislocations. The temperature independence for 1 mK $<$ \textit{T}
$<$ \textit{T$_S$} (see Fig.~\ref{fig:one}) suggests that below the
saturation point the dislocations are completely pinned. It is
surprising that even just 1 ppb (or less \cite{1ppb}) of impurities
can immobilize the dislocations. As a sample of a fixed
\textit{x}$_3$ is warmed above \textit{T$_S$} the continual
evaporation of $^3$He atoms from the dislocation network softens the
solid, which may concomitantly destroy NCRI. A dramatic increase at
low temperature of the shear modulus in solid $^4$He has in fact
been observed recently \cite{beamish3}. The anomaly exhibits the
same qualitative (and perhaps quantitative) dependencies on
temperature, $^3$He concentration (for \textit{x}$_3$ = 1 ppb, 85
ppb, and 300 ppb), and stress amplitude ($\propto$
\textit{v$_{RIM}$} in the TO experiments) as NCRI. We are further
investigating the connection between these experiments.

In conclusion, we found that the \textit{x}$_3$ dependence of the
characteristic temperatures of NCRI are consistent with the binding
of $^3$He atoms to dislocations. The absence of any temperature
dependence below $\sim$28 mK in isotopically-pure samples suggests
that the most likely phase transition point lays between
\textit{T$_S$} and \textit{T$_O$}.

\begin{acknowledgments}
We thank P. W. Anderson and J. R. Beamish for their advice. E. K.
and M. H. W. C. also acknowledge the illuminating discussions at
workshops held at the Kavli Institute of Theoretical Physics at UCSB
(2006), the Aspen Center for Physics (2006), Keio University (2007),
and the Outing Lodge sponsored by the Pacific Institute of
Theoretical Physics (2007). The work at PSU was supported by USA NSF
grants DMR-0207071 and DMR-0706339. The State of Florida and USA NSF
DMR-9527035 funded research carried out at UF.
\end{acknowledgments}


\begin{thebibliography}{25}
\bibitem{KC1} E. Kim and M. H. W. Chan, Nature \textbf{427}, 225 (2004).
\bibitem{KC2} E. Kim and M. H. W. Chan, Science \textbf{305}, 1941
(2004).
\bibitem{KC4} E. Kim and M. H. W. Chan, Phys. Rev. Lett. \textbf{97}, 115302 (2006).
%\bibitem{3Dsf} J. Wilks, \textit{The Properties of Liquid and Solid Helium} (Clarendon Press, Oxford, 1967).
%\bibitem{2Dsf} G. A. Csathy and M. H. W. Chan, Phys. Rev. Lett. \textbf{87}, 045301 (2001).
\bibitem{RR1} A. S. C. Rittner and J. D. Reppy, Phys. Rev. Lett. \textbf{97}, 165301 (2006).
\bibitem{RR2} A. S. C. Rittner and J. D. Reppy, Phys. Rev. Lett \textbf{98}, 175302 (2007).
\bibitem{HK} Y. Aoki, J. C. Graves, and H. Kojima, Phys. Rev. Lett. \textbf{99}, 015301 (2007).
\bibitem{MK} A. Penzev, Y. Yasuta, and M. Kubota, J. Low Temp. Phys. \textbf{148}, 677 (2007).
\bibitem{KS} M. Kondo, S. Takada, Y. Shimbayama, K. Shirahama, \textbf{148}, 695 (2007).
\bibitem{clark} A. C. Clark, J. T. West, and M. H. W. Chan, Phys. Rev. Lett. \textbf{99}, 135302 (2007).
\bibitem{KC3} E. Kim and M. H. W. Chan, J. Low Temp. Phys. \textbf{138}, 859 (2005).
\bibitem{1ppb} See reports BM-RI-8054, BM-RI-9010, and
PB-86-205309/XAB.
\bibitem{UHP} Additional numbers and references can be found at http://www.grisda.org/origins/25055.htm.
\bibitem{thesis} E. Kim, Ph.D. Thesis, The Pennsylvania State
University, 2004.
\bibitem{ganshin} A. N. Gan'shin \textit{et al.}, Low Temp. Phys. \textbf{26}, 869 (2000).
\bibitem{goodkind} J. M. Goodkind, AIP Conf. Proc. \textbf{850}, 329 (2006).
\bibitem{xi} X. Lin, A. C. Clark, and M. H. W. Chan, accepted for publication in Nature (London) (2007).
\bibitem{EB} D. O. Edwards and S. Balibar, Phys. Rev. B \textbf{39}, 4083 (1989).
\bibitem{anderson} P. W. Anderson, Nature Phys. \textbf{3}, 160 (2007).
\bibitem{umassclark} M. Boninsegni, A. B. Kuklov, L. Pollet, N. V. Prokof'ev, B. V. Svistunov, and M. Troyer, Phys. Rev. Lett. \textbf{99},
035301 (2007); A.C. Clark and M. H. W. Chan, to be published (2007).
\bibitem{iwasa} I. Iwasa and H. Suzuki, J. Phys. Soc. Jpn. \textbf{49}, 1722 (1980).
\bibitem{PBD} M. A. Paalanen, D. J. Bishop, and H. W. Dail, Phys. Rev. Lett. \textbf{46}, 664 (1981).
\bibitem{wanner} R. Wanner, I. Iwasa, and S. Wales, Solid State Commun. \textbf{18}, 853
(1976).
\bibitem{iwasa'} I. Iwasa, K. Araki, and H. Suzuki, J. Phys.
Soc. Jpn. \textbf{46}, 1119 (1979).
\bibitem{knorr} D. Wallacher, M. Rheinstaedter, T. Hansen, and K. Knorr, J. Low Temp. Phys. \textbf{138}, 1013 (2005).
\bibitem{beamish3} J. Day and J. R. Beamish, accepted for publication in Nature (London) (2007); arXiv:0709.4666v1 (2007).
\end{thebibliography}
\end{document}